\documentclass[twocolumn,epsf,prb,amsmath,amssymb,showpacs]{revtex4}
\usepackage{graphicx}
\usepackage{amsmath}

\begin{document}
\title{Landau levels in asymmetric graphene trilayers}

\author{J. M. Pereira Jr.$^1$, S. H. R. Sena$^{1}$, F. M. Peeters$^{1,2}$, and G. A. Farias$^1$}
\address{$^1$Departamento de F\'{\i}sica, Universidade Federal do Cear\'a,
Fortaleza, Cear\'a, 60455-760, Brazil.\\
$^2$Department of Physics, University of Antwerp,
Groenenborgerlaan 171, B-2020 Antwerpen, Belgium.}

\begin{abstract}
The electronic spectrum of three coupled graphene layers (graphene trilayers) is investigated in the
presence of an external magnetic field. We obtain analytical expressions for the Landau level spectrum for
both the ABA and ABC - type of stacking, which exhibit very different dependence on the magnetic field. 
The effect of layer asymmetry and of external gate voltages can
strongly influence the properties of the system.

\end{abstract}

 \maketitle 

\section{Introduction}

The extraordinary
level of interest in the study of single layer graphene has led to
the prediction and observation of several unusual phenomena not
found in other low-dimensional systems. These new properties are
mainly a consequence of the chiral and massless character of the
quasiparticles in graphene \cite{Review}. This new material is not
only expected to lead to several technological applications, but
has also helped shed light on relativistic quantum effects, such
as Klein tunneling \cite{Katsnelson,klein-review} and
zitterbewegung. There is currently a search for experimental
methods for producing high quality samples of graphene in large
quantities. However, the current experimental techniques can also
create carbon structures with two or more layers. It has already
been recognized that graphene bilayers (i.e. two coupled layers of
graphene) can display interesting new properties that are distinct
from those of single-layers \cite{McCann} and that can also be
eventually harnessed for the development of devices. One important
aspect of these structures is the fact that, in comparison with
the in-plane interactions, the comparatively weaker interlayer
coupling can yet exert a significant influence on the carrier
spectrum. Thus, whereas the electronic dispersion at the vicinity
of the Fermi energy in single layer graphene is linear, in bilayer
graphene it displays an approximately parabolic shape with the
appearance of higher energy bands. Moreover, in striking contrast
with single layers, the electronic spectrum of bilayer graphene
has been shown to develop a gap in the presence of an external
electric field. Thus, the interlayer coupling in stacked layers of
graphene gives rise to a rich set of properties that are not found
in monolayers, and can be expected to be of particular
significance in structures with three or more layers.

In this work we investigate the properties of three coupled layers of
graphene, i.e. trilayer graphene (TLG) in the presence of an
external magnetic field perpendicular to the plane of the layers.
The properties of TLG in the absence of a magnetic field have been
considered in the literature within a tight-binding model (see,
e.g. \cite{Bart,Artak1}), as well as through first principles
calculations \cite{Aoki}. The effect of an external magnetic field
was calculated by means of an approximation based on the mapping
of stacked graphene layers to an 1D tight-binding chain by
Guinea {\it et al.} \cite{Guinea}. 
Recent experimental studies have investigated the Landau level spectrum \cite{Thiti,Kumar}
and the magnetoconductance of TLG \cite{Bao,Bao2,Liu}.
These results showed that
one important aspect of TLG is the fact that the energy bands at
the vicinity of the Fermi energy are very sensitive to the
particular type of stacking of the layers. The two more relevant
stacking rules are the rhombohedral, or ABC stacking, and the
Bernal, or ABA stacking. In each case, the relative positions of
the lowermost layer (C in one case and A in the other) helps
dictate the possible symmetries of the subsequent wavefunctions
associated with each layer. The type of stacking is also relevant
to the properties of the Landau levels of the TLG, and it has recently
been reported \cite{Lui} that 15\% of exfoliated TLG has rhombohedral (ABC)
stacking. The goal of the
present paper is to present analytical results for the
spectrum of TLG in a magnetic field considering different potentials
in each layer. In order to
do that we perform a direct diagonalization of the six-band
continuum model and obtain analytical expressions for  
the Landau level spectrum as function of
magnetic field and the potentials at each layer, for both the ABC
and ABA stackings. In particular, we calculate the TLG spectrum in
the presence of electric fields that break the layer symmetry.
It has recently been shown that these different potentials can lead
to the opening of a gap in the TLG spectrum \cite{Artak1,Artak2,Koshino1,Heinz}. In the present paper
we focus on the modifications of the Landau levels as function of the
layer potential.

The paper is organized as follows: in section II we present the
model and solve the resulting system of equations for the ABA stacking,
followed by section III in which we present the model and obtain solutions
for the ABC stacking. Section IV shows and discusses the calculated
Landau level spectra and finally, the results are summarized in
the conclusions.

\begin{figure}
\centering
\includegraphics[width=8.5 cm]{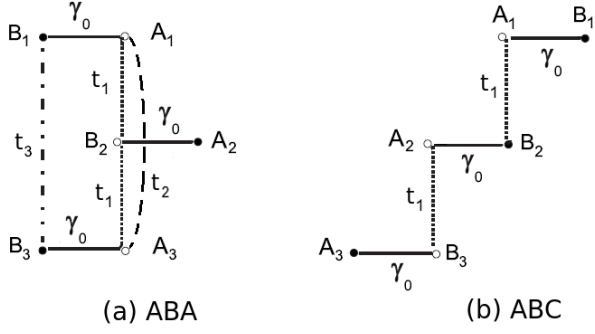}
\caption{ Diagramatic scheme of couplings in graphene trilayers
for ABC (a) and ABA (b) stackings.} \label{fig1}
\end{figure}

\section{ABA Stacking}

Let us consider a system consisting of three coupled graphene
layers, in the context of the tight-binding model. We assume nearest
neighbor hopping between sites within each layer, described by the
coupling parameter $t_0$. In the continuum approximation,
this parameter determines the magnitude of the Fermi velocity $v_F
= t_0 a \hbar^{-1} \sqrt{3}/2 \approx 10^6$ m$/$s. The
nearest neighbor interlayer coupling scheme is $A_1 - B_2  - A_3$,
with coupling parameter $t_1$ (see Fig. 1). In order to capture
some of the finer details of the spectrum, we also include a
remote coupling term $t_2$ between sites of sublattices $A_1$ and
$A_3$ and $t_3$ between $B_1$ and $B_3$. The Hamiltonian is given
as
\begin{equation}
{\mathcal H}=
\begin{pmatrix}
  U_1+U_0 & v_F \pi^\dagger & t_1 & 0 & t_2 & 0 \\
  v_F \pi & U_1 & 0 & 0 & 0 & t_3 \\
  t_1 & 0 & U_2+U_0 & v_F \pi & t_1 & 0 \\
  0 & 0 & v_F \pi^\dagger & U_2 & 0 & 0 \\
  t_2 & 0 & t_1 & 0 & U_3+U_0 & v_F \pi^\dagger \\
  0 & t_3 & 0 & 0 & v_F \pi & U_3
\end{pmatrix}
\end{equation}
where $\pi = p_x+ip_y$, with $p_{x,y}$ being the components of the in-plane momentum;
$U_{1,2,3}$ is the potential in each layer, respectively, $U_0$ is the onsite energy at
sublattices $A_1$, $B_2$ and $A_3$,
and we defined the eigenstates as $\Psi = [\psi_{A1}, \quad
i\psi_{B1}, \quad\psi_{B2}, \quad i\psi_{A2}, \quad\psi_{A3}, \quad
i\psi_{B3}]^T$. In the presence of a uniform magnetic field in the
z direction, with the gauge $\vec{A} = (0,Bx,0)$
and, for a given sublattice $L$, $\psi_L (y)= \phi_L e^{ik_y y}$, one obtains the following
system of equations:
\begin{subequations}
\begin{eqnarray}
&&{\mathcal A}^+\phi_{B1}+ t_1'\phi_{B2}+ t_2'\phi_{A3} =(\epsilon-u_1-u_0)\phi_{A1},\\
&&\cr && {\mathcal A}^-\phi_{A1} +t_3'\phi_{B3}= -(\epsilon-u_1)\phi_{B1},\\
&&\cr && {\mathcal A}^-\phi_{A2} + t_1'\phi_{A1} + t_1'\phi_{A3} = (\epsilon-u_2-u_0)\phi_{B2},\\
&&\cr && {\mathcal A}^+\phi_{B2} = -(\epsilon-u_2)\phi_{A2},\\
&&\cr && {\mathcal A}^+\phi_{B3} + t_1'\phi_{B2}+ t_2'\phi_{A1} = (\epsilon-u_3-u_0)\phi_{A3},\\
&&\cr && {\mathcal A}^-\phi_{A3} +t_3'\phi_{B1} = -(\epsilon-u_3)\phi_{B3},
\end{eqnarray}
\end{subequations}
where $\epsilon = E/\hbar v_F$, $u_i = U_i/\hbar v_F$, $t_i' =
t_i/\hbar v_F$ and $\beta = eB/\hbar v_F$ and we defined the operators
\begin{equation}
{\mathcal A}^{\pm} = \frac{d}{d x} \pm (k_y - \beta x),
\end{equation}
which obey the commutation relation $[{\mathcal A}^{+},{\mathcal A}^{-}] = 2\beta$.

For $U_1 = U_2 = U_3 = U$ the system can be easily solved by
making use of its reflection symmetry. Thus, we can define symmetric
and antisymmetric combinations of the spinor components. For the
antisymmetric case we obtain
$
\phi_G \equiv \frac{1}{\sqrt{2}}(\phi_{A1}-\phi_{A3}),$ and $
\phi_H \equiv \frac{1}{\sqrt{2}}(\phi_{B1}-\phi_{B3}).
$
That leads to the following pair of coupled equations
\begin{subequations}
\begin{eqnarray}
&&{\mathcal A}^-\phi_{G} = -(\epsilon-u-t_3')\phi_{H},\\ &&\cr && {\mathcal A}^+\phi_{H} =
(\epsilon-u+t'_2-u_0)\phi_{G}.
\end{eqnarray}
\end{subequations}
For the sake of convenience, let us now define the operator
\begin{equation}
Z \equiv  {\mathcal A}^-{\mathcal A}^+ =  \frac{d^2}{d x^2} - (k_y - \beta
x)^2-\beta .
\end{equation}
We can now decouple the equations to obtain
\begin{equation}
(Z + 2\beta )\phi_{G} = -[(\epsilon-u')^2-(\delta u)^2]\phi_{G},
\end{equation}
which corresponds to the equation that gives the spectrum for a {\it
single} graphene layer under an effective electrostatic potential
$U' = U + (t_3-t_2+U_0)/2$ as well as a finite gap term given by
$\delta U = (U_0 - t_2 - t_3)/2$. Thus, the Landau level spectrum in this case is
given by $\epsilon = u' \pm \sqrt{2\beta n + (\delta u)^2}$.
From Ref. [15] 
we have $t_2 = 0.04$ eV, $t_3 = -0.02$ eV and $U_0 = 0.05$ eV. Hence, the additional
effective potential leads to an energy shift of $-5.0$ meV, and $\delta U = 15$ meV.
Thus, the effect of the remote coupling terms $t_2$ 
and $t_3$, as well as the on-site energy term $U_0$ is only to introduce a small shift of the dispersion
branches and to generate a small gap in the energy spectrum. 

For the symmetric case we have
$
\phi_C \equiv \frac{1}{\sqrt{2}}(\phi_{A1}+\phi_{A3})$ and $
\phi_D \equiv \frac{1}{\sqrt{2}}(\phi_{B1}+\phi_{B3}).
$
The equations become
\begin{subequations}
\begin{eqnarray}
&&{\mathcal A}^+\phi_{D}+\sqrt{2}t_1'\phi_{B2} = (\epsilon-u-t_2'-u_0)\phi_{C},\\
&&\cr && {\mathcal A}^-\phi_{C} = -(\epsilon-u+t_3')\phi_{D},
\end{eqnarray}
and
\begin{eqnarray}
&&{\mathcal A}^-\phi_{A2}+\sqrt{2}t_1'\phi_{C} = (\epsilon-u-u_0)\phi_{B2},\\
&&\cr && {\mathcal A}^+\phi_{B2} =-(\epsilon-u)\phi_{A2},
\end{eqnarray}
\end{subequations}
These equations can be decoupled, resulting in the fourth-order differential equation
\begin{equation}
\{Z^2+\lambda_1 Z - \lambda_2 \} \phi_C = 0,
\end{equation}
where $\lambda_1 = (\epsilon-u-u_0)(\epsilon-u)+(\epsilon-u+t_3')(\epsilon-u-u_0-t_2')+2\beta$, and 
$\lambda_2 = -(\epsilon-u-u_0)(\epsilon-u)(\epsilon-u+t_3')(\epsilon-u-u_0-t_2')+2(t_1')^2(\epsilon-u)(\epsilon-u+t_3')$.
This equation is similar to the one describing bilayer
graphene. A second-order equation can be obtained by calculating the roots of
the second-order equation as
\begin{equation}
\{Z-z_+\} \{Z-z_-\}\phi_C = 0,
\end{equation}
with
\begin{equation}
z_\pm=-\frac{\lambda_1}{2}\pm\sqrt{\Big( \frac{\lambda_1}{2}\Big)^2+\lambda_2},
\end{equation}
where we set $k_y =0$, since this term only introduces a shift of the wavefunction
In particular, for $u_0$, $t_{2,3} = 0$ the equations yield results that
are identical to those of a gapless single-layer and bilayer graphene, i.e. the Landau levels
are found as the solutions of $2\beta(n+1) = z_\pm$. As in the previous case, 
the addition of remote coupling terms introduces a gap in the spectrum.

A more realistic description of TLG structures should take into account asymmetries between the different
layers, which can be brought about by the interaction with a substrate or by gating. In order to 
assess the effect of layer symmetry breaking in the spectrum, let us now consider the case $U_1 \neq U_2 \neq U_3$.
In addition, we now consider $t_2 = t_3 = U_0 = 0$, since we assume that the shifts caused by the layer potentials
are more significant than the effect of these terms. 
A simple substitution allows us to write
\begin{eqnarray}
&&[{\mathcal A}^+{\mathcal A}^- + (\epsilon-u_1)^2]\phi_{A1}=t_1'(\epsilon-u_1)\phi_{B2},\cr
&&\cr
&&[{\mathcal A}^+{\mathcal A}^- + (\epsilon-u_3)^2]\phi_{A3}=t_1'(\epsilon-u_3)\phi_{B2}.
\end{eqnarray}
In addition, we also have
\begin{equation}
[{\mathcal A}^-{\mathcal A}^+ + (\epsilon-u_2)^2]\phi_{B2}= t_1'(\epsilon-u_2)(\phi_{A1}+\phi_{A3}). 
\end{equation}
As in the previous case, we introduce symmetric and antisymmetric combinations of wavefunctions, and
let us also define $\Delta = (u_1-u_3)/2$, $s = (u_1+u_3)/2$ and $\delta_j = \epsilon - u_j$, $j=1,2,3$, in order to
simplify the notation. 
Thus, after some algebra, we can obtain the following 6th-order differential equation for $\phi_{B2}$ as
\begin{eqnarray}
\Big\{[{\mathcal A}^-{\mathcal A}^+ &+& \delta_1^2 + 2\beta][{\mathcal A}^-{\mathcal A}^+ + {\delta_2}^2][{\mathcal A}^-{\mathcal A}^+ + {\delta_3}^2 + 2\beta]\cr
&&\cr
&&-t_1'^2\delta_2(\epsilon - s)[{\mathcal A}^-{\mathcal A}^+ + {\delta_1}^2 + 2\beta]\cr
&&\cr
&&-t_1'^2\delta_2(\epsilon - s)[{\mathcal A}^-{\mathcal A}^+ + {\delta_3}^2 + 2\beta] \cr
&&\cr
&&+4t_1'^2\Delta^2\delta_2(\epsilon - s)\Big\} \phi_{B2} = 0. 
\end{eqnarray}
It is seen that for $U_1=U_2=U_3$ (i.e. $\delta_1=\delta_2=\delta_3$, $\Delta = 0$), we recover the previous solutions.
One can rewrite Eq. (22) as
\begin{equation}
[ Z^3 + \alpha_1 Z^2 + \alpha_2 Z + \alpha_3 ]\phi_{B2} = 0 .
\end{equation}
with the $Z$ operator defined above and
\begin{subequations}
\begin{eqnarray}
\alpha_1 \equiv &&\delta_1^2 + \delta_2^2 + \delta_3^2 + 4\beta,\\
&&\cr
\alpha_2 \equiv &&(\delta_1^2+2\beta)(\delta_3^2+2\beta)+(\delta_1^2+2\beta)\delta_2^2 + (\delta_3^2+2\beta)\delta_2^2 \cr
&&\cr
&&- t_1'^2\delta_2(\delta_1+\delta_3),\\
&&\cr
\alpha_3 \equiv &&(\delta_1^2+2\beta)\delta_2^2(\delta_3^2+2\beta)-2\beta t_1'^2\delta_2(\delta_1+\delta_3)\cr
&&\cr
&& -t_1'^2\delta_1\delta_2\delta_3(\delta_1 + \delta_3),
\end{eqnarray}
\end{subequations}
This equation can be written as
\begin{equation}
\{Z-Z_1\}\{Z-Z_2\}\{Z-Z_3\}\phi_{B2} = 0,
\end{equation}
where $Z_j$, $j=1,2,3$ are the three roots of the cubic equation, Eq. (14). Therefore, 
the spinor component $\phi_{B2}$ is a solution of
\begin{equation}
-\frac{d^2\phi_{B2}}{dx^2}+(k_y-\beta x)^2\phi_{B2}=-(Z_j+ \beta)\phi_{B2}.
\end{equation}
For zero magnetic field, this equation allows us to obtain plane wave solutions for each dispersion branch. The dispersion relation
can be obtained by setting $Z_j = -k^2$. It can be immediately seen that the energy gap at $k=0$ can be found by solving the
equation $\alpha_3 = 0$. For finite
magnetic fields, the solutions are expressed in terms of Hermite polynomials.
Therefore, for the Landau levels we obtain the relation $Z_j = -2\beta (n+1)$. Thus, the energies are found by solving the algebraic equation
\begin{equation}
-[2\beta (n+1)]^3+\alpha_1 [2\beta (n+1)]^2 - \alpha_2 [2\beta (n+1)] + \alpha_3 = 0. 
\end{equation}
It is evident that for $U_1=U_2=U_3$, we have $\Delta = 0$ and the last term of Eq. (13) vanishes. The spectrum should then consist of a superposition of the 
spectra of single layer graphene and bilayer graphene.

\section{ABC Stacking}

Let us consider three coupled graphene layers in the ABC
stacking configuration. For the sake of simplicity, let us retain only the nearest-neighbor coupling terms.
In this case, the Hamiltonian can be written as
\begin{equation}
{\mathcal H}=
\begin{pmatrix}
  U_1 & v_F \pi^\dagger & t & 0 & 0 & 0 \\
  v_F \pi & U_1 & 0 & 0 & 0 & 0 \\
  t & 0 & U_2 & v_F \pi & 0 & 0 \\
  0 & 0 & v_F \pi^\dagger & U_2 & 0 & t \\
  0 & 0 & 0 & 0 & U_3 & v_F \pi^\dagger \\
  0 & 0 & 0 & t & v_F \pi & U_3
\end{pmatrix}
\end{equation}
where $U_{1,2,3}$ is the potential in each layer, respectively, and we defined the eigenstates as before.

Thus, one can obtain the following system of equations:
\begin{subequations}
\begin{eqnarray}
&&{\mathcal A}^+\phi_{B1}+ t'\phi_{B2}
= (\epsilon-u_1)\phi_{A1},\\ &&\cr &&
{\mathcal A}^-\phi_{A1} = -(\epsilon-u_1)\phi_{B1},\\ &&\cr &&
{\mathcal A}^-\phi_{A2} + t'\phi_{A1} =
(\epsilon-u_2)\phi_{B2},\\ &&\cr &&
{\mathcal A}^+\phi_{B2} - t'\phi_{B3}=
-(\epsilon-u_2)\phi_{A2},\\ &&\cr &&
{\mathcal A}^+\phi_{B3} =
(\epsilon-u_3)\phi_{A3},\\ &&\cr &&
{\mathcal A}^-\phi_{A3} - t'\phi_{A2} =
-(\epsilon-u_3)\phi_{B3},
\end{eqnarray}
\end{subequations}
where $\epsilon = E/\hbar v_F$, $u_i = U_i/\hbar v_F$, $t' = t/\hbar v_F$ and $\beta = eB/\hbar v_F$.
In order to decouple these equations, let us first obtain $\phi_{B1}$ and $\phi_{A3}$ in terms of $\phi_{A1}$ and $\phi_{B3}$ from the second and
fifth equations as
\begin{equation}
\phi_{B1} = -\frac{1}{(\epsilon - u_1)}{\mathcal A}^-\phi_{A1}, \qquad \phi_{A3} =\frac{1}{(\epsilon - u_3)}{\mathcal A}^+\phi_{B3},
\end{equation}
and substitute these expressions in the first and sixth equations, respectively, to give
\begin{subequations}
\begin{eqnarray}
&&{\mathcal A}^+{\mathcal A}^-\phi_{A1} - t'(\epsilon - u_1)\phi_{B2} = -(\epsilon - u_1)^2\phi_{A1},\quad \\ &&\cr &&
{\mathcal A}^-{\mathcal A}^+\phi_{B3} - t'(\epsilon - u_3)\phi_{A2} = -(\epsilon - u_3)^2\phi_{B3}.\quad
\end{eqnarray}
\end{subequations}
Equations (22a) and (22b) allow us to obtain $\phi_{B2}$ and $\phi_{A2}$ in terms of $\phi_{A1}$ and $\phi_{B3}$, respectively.
Thus, by substituting them in Eqs. (27) and (28), respectively, and after some tedious algebra, one can obtain a 6th order differential equation as
\begin{eqnarray}
\Bigl\{&&[{\mathcal A}^-{\mathcal A}^+ + \delta_1^2 +2\beta][{\mathcal A}^-{\mathcal A}^+ + \delta_2^2][{\mathcal A}^-{\mathcal A}^+ + \delta_3^2 -2\beta] \cr
&&-t'^2\delta_2\delta_3 [{\mathcal A}^-{\mathcal A}^+ + \delta_1^2 +2\beta]\cr
&&\cr
&&-t'^2\delta_1\delta_2 [{\mathcal A}^-{\mathcal A}^+ + \delta_3^2 -2\beta] + t'^4 \delta_1\delta_3\Bigr\}\phi_{A1} =0.
\end{eqnarray}
It is interesting to compare Eqs. (13) and (23). The former remains invariant if one switches the
potentials in layers $1$ and $3$. Equation (23), on the other hand, is found to be invariant
under an interchange of potentials between the topmost and lowest layers together with a reversal of
the magnetic field. This reflects the different symmetries of each stacking of TLG.

As before, we can obtain the Landau level spectrum by rewritting Eq. (23) as
\begin{equation}
[Z^3+\gamma_1 Z^2 + \gamma_2 Z + \gamma_3]\phi_{A1} = 0, 
\end{equation}
with $Z \equiv {\mathcal A}^-{\mathcal A}^+$, and
\begin{subequations}
\begin{eqnarray}
\gamma_1 \equiv &&\delta_1^2 + \delta_2^2 + \delta_3^2,\\
&&\cr
\gamma_2 \equiv &&(\delta_1^2+2\beta)(\delta_3^2-2\beta)+(\delta_1^2+2\beta)\delta_2^2 + (\delta_3^2-2\beta)\delta_2^2\cr
&&\cr
&&- (t')^2 (\delta_1 + \delta_3)\delta_2,\\
&&\cr
\gamma_3 \equiv &&-(t')^2\delta_2 [\delta_3(\delta_1^2+ 2\beta ) + \delta_1(\delta_3^2 - 2\beta)] + (t')^4 \delta_1\delta_3\cr
&&\cr 
&& +(\delta_1\delta_2\delta_3)^2+2\beta(\delta_3^2-\delta_1^2)\delta_2^2-4(\beta \delta_2)^2,
\end{eqnarray}
\end{subequations}
This equation can be written as
\begin{equation}
\{Z-Z_1\}\{Z-Z_2\}\{Z-Z_3\}\phi_{A1} = 0,
\end{equation}
where $Z_j$, $j=1,2,3$ are the three roots of the cubic equation, Eq. (24). 
Therefore the spinor component $\phi_{A1}$ is found as a solution of
\begin{equation}
-\frac{d^2\phi_{A1}}{dx^2}+(k_y-\beta x)^2\phi_{A1}=-(Z_j+ \beta)\phi_{A1}.
\end{equation}
For the particular case of $U_1=U_2=U_3=0$ and zero magnetic field, we can obtain plane wave solutions by setting $Z_j = -k^2$, where
$k$ is the in-plane wavevector. Thus, Eq. (24) can be rewritten as
\begin{equation}
\epsilon^6 - (3k^2 + 2t'^2)\epsilon^4 + (3k^4 - 2k^2t'^2+ t'^4)\epsilon^2 - k^6 = 0.
\end{equation}
Let us now consider the low-energy limit $\epsilon << t'$. That allows us to neglect the higher-order
powers of $\epsilon$ to obtain
\begin{equation}
\epsilon \approx \frac{k^3}{t'^2} \frac{1}{\sqrt{1-2k^2/t'^2+3k^4/t'^4}}.
\end{equation}
Thus, for small wavevectors, the dispersion relation increases with the third power of $k$.

The Landau levels can be obtained using the relation $Z_j = -2\beta (n+1)$, which leads to the algebraic equation
\begin{equation}
-[2\beta (n+1)]^3+\gamma_1 [2\beta (n+1)]^2 - \gamma_2 [2\beta (n+1)] + \gamma_3 = 0. 
\end{equation}
As seen above, for zero potential in each layer, we can find a simpler algebraic relation for the energy, namely
\begin{eqnarray}
&&\epsilon^6 - [6\beta (n+1) + 2t'^2]\epsilon^4 \cr
&&\cr
&&+[12\beta^2 (n+1)^2 - 4\beta (n+1)t'^2+ t'^4 - 4\beta^2]\epsilon^2 \cr
&&\cr
&&- 8\beta^3 (n+1)^3+8\beta^3(n+1) = 0. 
\end{eqnarray} 
For $\epsilon << t$ we can then obtain
\begin{equation}
\epsilon \approx \pm\frac{(2\beta)^{3/2}}{t'^2}\sqrt{n(n+1)(n+2)} F(\beta,n), 
\end{equation}
where 
\begin{equation}
F(\beta,n) = \Big[ 1-4\frac{\beta}{t'^2}(n+1)-4\frac{\beta^2}{t'^4}+12\frac{\beta^2}{t'^4}(n+1)^2 \Big]^{-1/2}.
\end{equation}
For small fields (i.e. $\beta << t'^2$), $F(\beta,n) \approx 1$.
Therefore, in the limit of low energies and small fields, the Landau levels should approximately depend on the magnetic field as $B^{3/2}$,
in agreement with the results of Ref. [9]. 

\section{Numerical Results}

\begin{figure}
\includegraphics*[height=7.0cm, width=8.0cm]{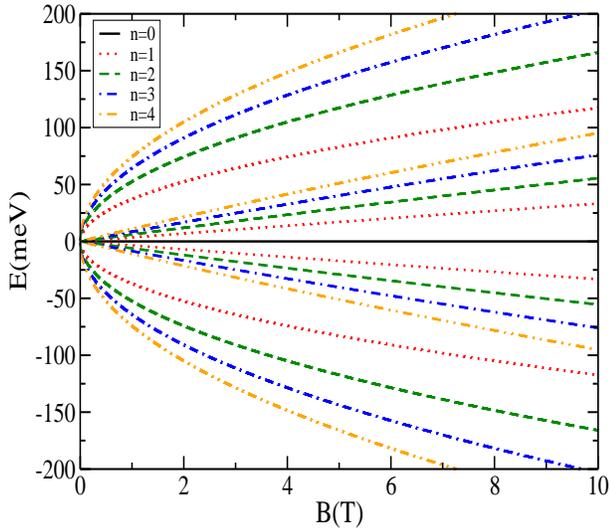}
\caption{
(Color online). The lowest Landau levels as function of magnetic field for ABA-stacked graphene trilayers calculated from Eq. (18), with $U_1=U_2=U_3=0$, 
for $n=0$ (black solid lines), $n=1$
(red dotted lines), $n=2$ (green dashed lines), $n=3$ (blue dot-dashed lines) and $4$ (yellow dot-dot-dashed lines).
}
\end{figure}

\begin{figure}
\includegraphics*[height=7.0cm, width=8.0cm]{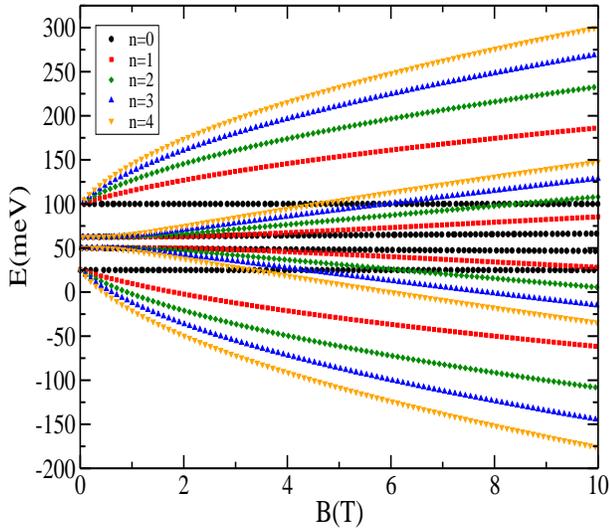}
\caption{
(Color online). Energy spectrum as function of magnetic field for ABA-stacked graphene trilayers, for $U_1=100$ meV, $U_2 = 50$ meV
and $U_3 = 25$ meV, $n=0$ (black dots), $n=1$
(red squares), $n=2$ (blue lozenges), $n=3$ (green triangles) and $4$ (yellow triangles). }
\end{figure}
\begin{figure}
\includegraphics*[height=7.0cm, width=8.0cm]{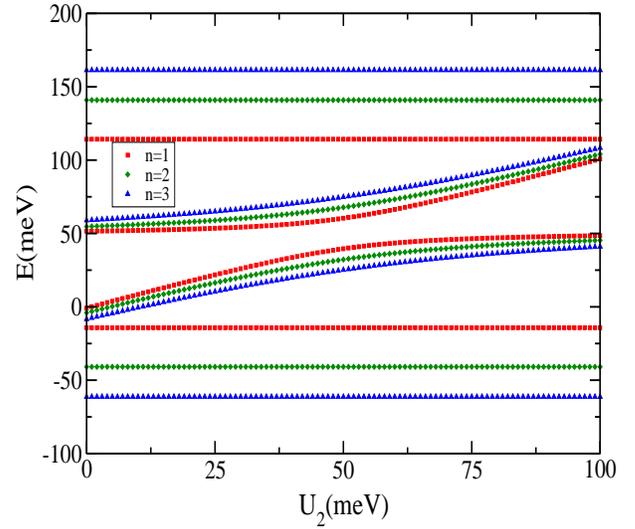}
\caption{
(Color online). Low-lying Landau levels as function of the potential in the inner layer for ABA-stacked graphene trilayers, for $n=1$ (red squares), $2$ (green lozenges) 
and $3$ (blue squares) for $B = 3$ T, $U_1 = U_3 = 50$ meV.}
\end{figure}

\begin{figure}
\includegraphics*[height=7.0cm, width=8.0cm]{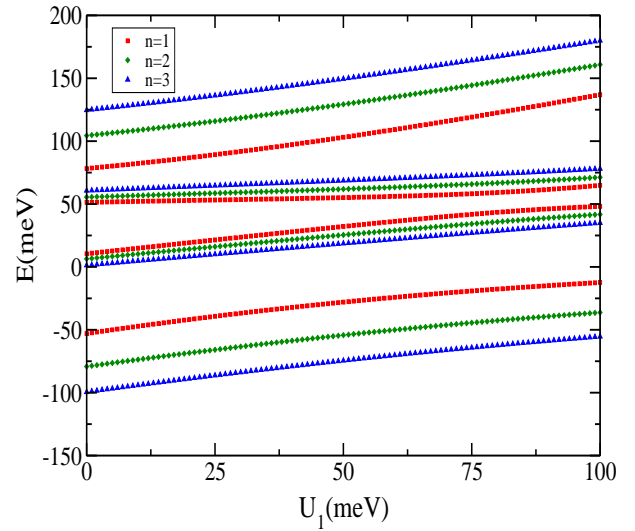}
\caption{
(Color online). Low-lying Landau levels as function of the potential in the uppermost layer for ABA-stacked graphene trilayers, for $n=1$ (red squares), $2$ (green lozenges) 
and $3$ (blue squares) for $B = 3$ T, $U_2 = 50$ meV, $U_3 = 25$ meV.}
\end{figure}

\begin{figure}
\includegraphics*[height=8.0cm, width=8.0cm]{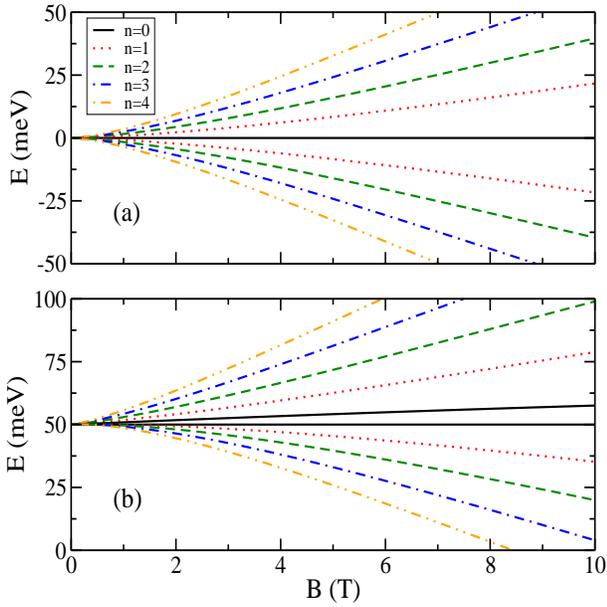} 
\caption{
(Color online). Landau level spectrum for the trilayer graphene for the ABC stacking, as function of magnetic field, with $U_1=U_2=U_3=0$ (a), and 
$U_1 = U_3 = 50$ meV, $U_2 = 100$ meV (b) for $n=0$ (black solid lines), $n=1$
(red dotted lines), $n=2$ (green dashed lines), $n=3$ (blue dot-dashed lines) and $4$ (yellow dot-dot-dashed lines).}
\end{figure}

\begin{figure}
\includegraphics*[height=6.5cm, width=8.0cm]{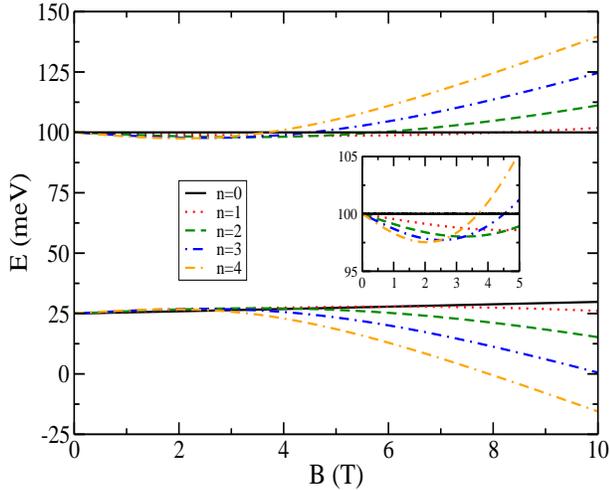}
\caption{
(Color online). Landau level spectrum for the trilayer graphene for the ABC stacking, as function of magnetic field, with $U_1 = 100$ meV, $U_2 = 50$ meV, and $U_3 = 25$ meV, 
for $n=0$ (black solid lines), $n=1$
(red dotted lines), $n=2$ (green dashed lines), $n=3$ (blue dot-dashed lines) and $4$ (yellow dot-dot-dashed lines).}
\end{figure}

\begin{figure}
\includegraphics*[height=6.5cm, width=8.0cm]{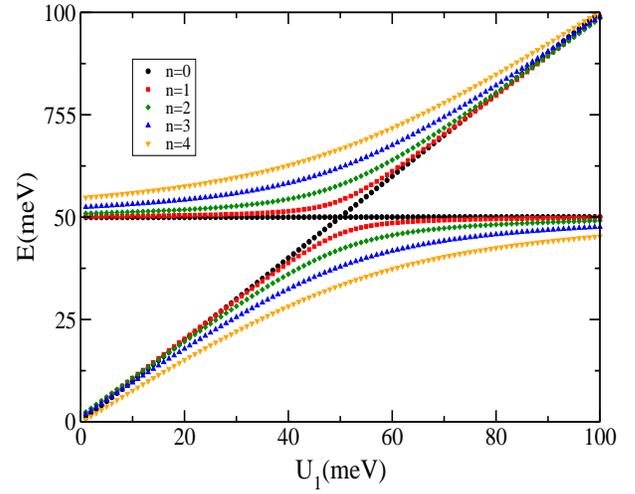}
\caption{
(Color online). Landau level spectrum for ABC-stacked trilayer graphene as function of $U_1$, for $B = 3$ T and
$U_2 = U_3 = 50$ meV, with $n=0$ (black dots), $n=1$
(red squares), $n=2$ (blue lozenges), $n=3$ (green triangles) and $4$ (yellow triangles).}
\end{figure}

\begin{figure}
\vspace*{1.0cm}
\includegraphics*[height=7.0cm, width=8.0cm]{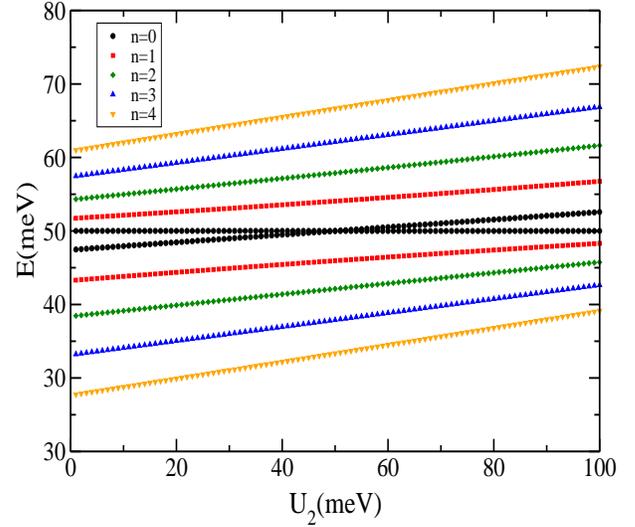}
\caption{
(Color online). Landau level spectrum for ABC-stacked trilayer graphene as function of $U_2$, for $B = 3$ T and
$U_1 = U_3 = 50$ meV, with $n=0$ (black dots), $n=1$
(red squares), $n=2$ (blue lozenges), $n=3$ (green triangles) and $4$ (yellow triangles).}
\end{figure}

Let us first consider the ABA case. Figure 2 shows the field dependence of the low-lying Landau levels calculated from Eq. (18), 
with $U_1=U_2=U_3=0$, for $n=0$ (black solid lines), $n=1$
(red dotted lines), $n=2$ (green dashed lines), $n=3$ (blue dot-dashed lines) and $4$ (yellow dot-dot-dashed lines). 
We find that for every value of $n$ there are two different types of low-energy branches: 1) those that depend linearly 
on the magnetic field (i.e. ``bilayer-like'' behavior), and 2) branches that display a
$B^{1/2}$ dependence (``monolayer-like'' branches). A third set of bilayer-like branches are found around $E = \pm t_1$, not shown in the
figure. 

The effect of a potential difference between the layers on the energy spectrum as function of magnetic field is shown in Fig. 3.
As in the previous case, the figure shows branches corresponding to $n=0 - 4$. The potentials in the different graphene layers are $U_1=100$ meV, $U_2 = 50$ meV
and $U_3 = 25$ meV. As seen from Eq. (18), for $B \rightarrow 0$ we have solutions corresponding to $E = U_j$, $j=1,2,3$ and $E = (U_1 + U_3)/2$.
Thus, we find that the ``monolayer-like'' branches are shifted creating a gap with magnitude $U_1 - U_3$, whereas for the ``bilayer-like'' states
a smaller gap opens with magnitude $(U_1+U_3)/2 - U_2$. One consequence of this difference is the appearance of level crossings as the magnetic field is
increased.

Figure 4 shows the low-energy Landau levels as function of the potential in the inner layer, for the ABA case, for $n=1$ (red squares), $2$ (green lozenges) 
and $3$ (blue squares) for $B = 3$ T, $U_1 = U_3 = 50$ meV. 
Notice that: 1) the lowest energy levels depend linearly on $U_2$ for small $U_2$ ($<< U_1 = U_3$) and for large
$U_2$ ($>> U_1 = U_3$) values, and exhibit an anticrossing behavior for $U_2 \approx U_1 = U_3$; 2) the higher energy states are very weakly affected by the bias.
A different behaviour is observed when one varies the potential at the uppermost layer ($U_1$), as seen in Fig. 5. In contrast with the previous case, the
bias is seen to cause a significant shift also on higher-energy Landau levels. 

The energy spectrum for the ABC case is shown in Fig. 6a, as function of magnetic field and with $U_1=U_2=U_3=0$, for $n = 0 - 4$.
For small B-values we see a doubly-degenerate branch with $E =0$, and a $B^{3/2}$-behavior for the remaining branches, which turns into a linear behavior at 
large $E$. In comparison with the previous case, the results in the ABC case
show the presence of pairs of branches at low energies, whereas in the ABA case one finds two sets of energy levels for each Landau index. That is caused by
the fact that, in the ABC case, the remaining four branches are found around $E = \pm t$, with $t \approx 400$ meV. 

Figure 6b shows results for an ABC TLG with $U_1 = U_3 = 50$ meV, whereas $U_2 = 100$ meV. In this case, the main effect of the potential difference is
the lifting of the degeneracy of the $n = 0$ state and a shift of the whole spectrum to lower energy with increasing magnetic field. 

In contrast, Fig. 7 shows the LL spectrum for $U_1 = 100$, $U_2 = 50$ and $ U_3 = 25$ meV. The inset shows an enlargement of the region around $E = 100$ meV. 
In this case, the bias creates an energy gap, which can be found by setting $\beta = 0$ in Eq. (30), which leads to solutions with $E = U_1$ and $E = U_2$.
Notice also the existence of level crossings, as well as the peculiar small magnetic field behavior where there is a reversal of the ordering of 
the Landau levels as compared to the regular high magnetic field behavior.

Results for the dependence of the energy spectrum on $U_1$ is shown in Fig. 8, for $B = 3$ T and
$U_2 = U_3 = 50$ meV. As seen, the degeneracy of the $n=0$ is lifted for $U_1 \neq U_2, U_3$. Moreover, when the magnitude of the potential in the 
uppermost layer is increased, the Landau levels tend to become degenerate. A quite distinct picture emerges if one varies instead the potential in
the middle layer ($U_2$), as shown in Fig. 10, for $B = 3$ T and $U_1 = U_3 = 50$ meV. In contrast with the previous results, the spectrum shows a 
linear dependence on the potential and there are no degeneracies for the different Landau indices. As in the previous figure, a single Landau level at $E = 50$ meV is
found to be unaffected by the bias.

\section{Conclusions}

In summary, we obtained exact analytical expressions for the Landau level spectra of 
trilayer graphene, within a model that took into account the layer asymmetry induced by 
different electrostatic potentials in each layer. The expressions were obtained for both the
Bernal (ABA) and rhombohedral (ABC) stackings, which were found to display quite distinct behaviors.
As shown in previous work, the Landau level spectrum for the ABA case in the absence of electrostatic
bias between the layers shows both a monolayer-like as well as bilayer-like character, indicated by
the different magnetic field dependence of the spectrum. The addition of a potential difference between
the layers shifts the spectrum and creates a tunable gap between the electron and hole states, the size of
this energy gap being different for the ``monolayer'' and the ``bilayer'' energy levels.
Level crossings between the ``monolayer'' and ``bilayer'' Landau levels are found for certain values
of the magnetic field. 

For the ABC case, the Landau levels have a magnetic field dependence which, in the
absence of bias, has a $B^{3/2}$-dependence for low energies.
For stronger magnetic fields the Landau levels exhibit a linear $B$-dependence. 
The introduction of electrostatic bias in the system lifts the degeneracy of the $n=0$ 
levels and creates a tunable gap. The results show also the existence of level crossings at small magnetic fields. 
This model can be refined by taking into account second-nearest neighbor
terms, as well as remote coupling between the lowest and uppermost layers. However, these additional terms
are expected not to influence the qualitative agreement behavior of the present results.

\section{Acknowledgements}
This work was supported by the Brazilian
Council for Research (CNPq), the Flemish Science Foundation
(FWO-Vl), the Belgian Science Policy (IAP) and the bilateral projects between Flanders and Brazil and
the CNPq and FWO-Vl.


\begin{thebibliography}{9}



\bibitem{Review}
A. H. Castro Neto, F. Guinea, N. M. R. Peres, K. S. Novoselov, and A. Geim,
Rev. Mod. Phys. {\bf 81}, 109 (2009).

\bibitem{Katsnelson}
M. I. Katsnelson, K. S. Novoselov, and A. K. Geim, Nature Physics
{\bf 2}, 620 (2006).

\bibitem{klein-review}
J. M. Pereira Jr., F. M. Peeters, A. Chaves, and G. A. Farias,
Semic. Sci. Tech. {\bf 25}, 033002 (2010).


\bibitem{milton1}
J. M. Pereira Jr., P. Vasilopoulos, and F. M. Peeters, Appl. Phys.
Lett. {\bf 90}, 132122 (2007).

\bibitem{McCann}
E. McCann and V. I. Fal'ko, Phys. Rev. Lett. {\bf 96}, 086805
(2006).

\bibitem{Bart}
B. Partoens, and F. M. Peeters, Phys. Rev. B {\bf 74}, 075404
(2006).

\bibitem{Artak1}
A. A. Avetisyan, B. Partoens, and F. M. Peeters, 
Phys. Rev. B {\bf 81}, 115432 (2010). 

\bibitem{Aoki}
M. Aoki, H. Amawashi, Solid State Commun. {\bf 142}, 123 (2007).

\bibitem{Guinea}
F. Guinea, A. H. Castro Neto, and N. M. R. Peres,
Phys. Rev. B {\bf 73}, 245426 (2006).

\bibitem{Thiti} T. Taychatanapat, K. Watanabe, T. Taniguchi, and P. Jarillo-Herrero, 
Nature Phys. (2011) 

\bibitem{Kumar} A. Kumar, W. Escoffier, J.M. Poumirol, C. Faugeras, D. P. Arovas, 
M. M. Fogler, F. Guinea, S. Roche, M. Goiran, and B. Raquet, arXiv:1104.1020v1

\bibitem{Bao} W. Bao, Z. Zhao, H. Zhang, G. Liu, P. Kratz, L. Jing, J. Velasco, Jr., D. Smirnov, and C. N. Lau, 
Phys. Rev. Lett. {\bf 105}, 246601 (2010).

\bibitem{Bao2} W. Bao, L. Jing, Y. Lee, J. Velasco Jr., P. Kratz, D. Tran, B. Standley, M. Aykol, S. B. Cronin, D. Smirnov, 
M. Koshino, E. McCann, M. Bockrath, and C.N. Lau, arXiv:1103.6088v1 

\bibitem{Liu} Y. Liu, S. Goolaup, C. Murapaka, W. S. Lew, and S. K. Wong,
ACS Nano. {\bf 4}, 7087 (2010).

\bibitem{Lui}
C. H. Lui, Z. Li, Z. Chen, P. V. Klimov, L. E. Brus, and T. F. Heinz,
Nano Lett. {\bf 11}, 164 (2011).

\bibitem{Artak2}
A. A. Avetisyan, B. Partoens, and F. M. Peeters,
Phys. Rev. B {\bf 80}, 195401 (2009).

\bibitem{Koshino1}
M. Koshino, and E. McCann,
Phys. Rev. B {\bf 79}, 125443 (2009).

\bibitem{Heinz}
C. H. Lui, Z. Li, K. F. Mak, E. Cappelluti, T. F. Heinz, arXiv:1105.4658v1



\end{thebibliography}
\end{document}